\documentclass[10pt,rmp,twocolumn,nofootinbib,amsmath,floatfix,aps]{revtex4-1}

\usepackage{graphicx}
\usepackage{epsfig}

\newcommand{\ldb}{\lambda_\textrm{dB}}

\begin{document}
\title{Colloquium: Quantum interference of clusters and molecules}

\author{Klaus Hornberger}
\affiliation{University of Duisburg-Essen\mbox{, }Faculty of Physics,
\\
Lotharstra{\ss}e 1-21\mbox{, }47048 Duisburg\mbox{, }Germany}
\author{Stefan Gerlich, Philipp Haslinger, Stefan Nimmrichter, and Markus Arndt}
\affiliation{University of Vienna\mbox{, }Vienna Center for Quantum Science and Technology  VCQ, 
\\
Faculty of Physics\mbox{, }Boltzmanngasse 5\mbox{, }1090 Vienna\mbox{, }Austria.}

\begin{abstract}
We review recent progress and future prospects of matter wave interferometry with complex organic molecules and inorganic clusters.
Three variants of a near-field interference effect, based on diffraction by material nanostructures, at optical phase gratings, and at ionizing laser fields are considered.
We discuss the theoretical concepts underlying these experiments and the experimental challenges. This includes optimizing interferometer designs as well as understanding the role of decoherence.
The high sensitivity of matter wave interference experiments to external perturbations is demonstrated to be useful for accurately measuring internal properties of delocalized nanoparticles.
We conclude by investigating the prospects for probing the quantum superposition principle in the limit of high particle mass and complexity.
\\[\baselineskip]\noindent
Published in: Reviews of Modern Physics 84 (2012) 157-173
\end{abstract}

\maketitle
\tableofcontents

\section{Introduction}  \label{Intro}
The wave nature of matter, most conspicuously revealed in interference studies such as the double slit experiment, is a paradigm of quantum mechanics. According to Richard Feynman, it even `contains the  only mystery' of quantum physics \cite{Feynman1965}. And indeed, the spatial delocalization of objects composed of hundreds of atoms, over extensions that exceed the particle size by orders of magnitude, clearly defies our common intuition. In spite of that,
such highly non-classical states have been created repeatedly in the laboratory and are used for fundamental science.

Recent matter wave experiments with nanometer-sized particles have opened a new field of research at the interface between the foundations of physics, quantum optics and physical  chemistry.
Here we focus mainly on near-field interference experiments, since they exhibit a number of advantages over the conceptually simpler far-field arrangements, if one is interested in massive and internally complex particles.
An overview of earlier experiments with more elementary particles can be obtained from the
comprehensive reviews on interference with electrons \cite{Tonomura1987,Batelaan2007,Hasselbach2010}, neutrons \cite{Rauch2000}, and atoms \cite{Berman1997, Cronin2009}.

There are two main motivations for probing the wave nature of complex particles. First, we will explain how it can be exploited to study the internal properties and dynamics of quantum delocalized particles. Quantum-assisted molecule metrology is now becoming a viable tool for molecular physics with prospects to outperform classical measurements in the near future.
Second,  nanoparticle interference is well suited for studying the quantum superposition principle in a mass regime that has not been accessible hitherto. We will show how recent studies can validate quantitatively the predictions made by decoherence theory, and we will argue that matter-wave experiments will set bounds to theories predicting a modification of the Schr\"odinger equation at the quantum-classical borderline.

We start in Sect.~\ref{Concepts} by discussing the general requirements for de Broglie wave interferometry and the advantages of near-field experiments.
Section~\ref{sec:interferometers} describes three recent near-field interferometer developments, their merits and drawbacks. The highly non-classical states in these instruments enabled studies of environmental decoherence, which are reviewed in Section~\ref{sec:decoherence}. While some types of interactions with the environment induce the emergence of classical behavior, others can be exploited to measure internal molecular properties, as explained in Section~\ref{Metrology}.

In Sect.~\ref{ExperimentalChallenges} we present some challenges encountered in the attempt to extend matter wave interferometry to ever more complex objects. This includes requirements for molecular beam sources, detection schemes and ways to cope with phase averaging.
A quantitative theoretical approach to describe all of these experiments is briefly introduced in Sect.~\ref{Theory}.
So far, there has been no experimental indication that the quantum superposition principle may fail at any mass or length scale. However, as discussed in Sect.~\ref{FundamentalLimits}, a number of modifications of the Schr\"odinger equation have been suggested, and in particular the model of continuous spontaneous localization may be the first alternative that can be put to a definitive test in matter wave experiments with very massive clusters.
We close in Sect.~\ref{Conclusions} with our conclusions and an outlook.

\section{From far-field to near-field interferometry}\label{Concepts}
From a conceptual point of view, far-field diffraction is by far the simplest and most palpable matter wave phenomenon: When a collimated and sufficiently slow particle beam impinges on a mask perforated by equidistant slits the particle density further downstream exhibits a fringe pattern, whose period is determined by the ratio $\ldb/d$ of the de Broglie wavelength $\ldb$ and the slit separation $d$. Such experiments have been implemented with electrons \cite{Jonsson1961}, neutrons \cite{Zeilinger1988}, and atoms \cite{Keith1988,Carnal1991,Shimizu1992}. More recently, also beams composed of cold helium clusters \cite{Schollkopf1994,Bruhl2004}, and hot fullerenes \cite{Arndt1999} have shown such far-field interference patterns,  in complete agreement with quantum expectations, including the subtle but significant role of van der Waals forces between the molecules and the material grating structures.

A number of alternative interferometer concepts have been studied also with diatomic molecules:
This includes Mach-Zehnder interferometry with Na$_2$, using three nanofabricated gratings \cite{Chapman1995b}, Ramsey-Bord{\'e} interference experiments, exploiting the near-resonant interaction with four running laser waves, with I$_2$ \cite{Borde1994} and K$_2$ \cite{Lisdat2000}, and the observation of the Poisson spot behind a circular obstacle with D$_2$  \cite{Reisinger2009}.
The scattering of fast H$_2$, as first studied by \textcite{Estermann1930}, has recently been extended to quantum reflection studies with He$_2$ by \textcite{Zhao2011}.

Most of these experiments operate in the far field, thus requiring a collimation of the molecular beam that is significantly narrower than the diffraction angle. This condition is the reason why it is difficult to extend these far-field schemes to
objects composed of, say, several hundred thousand atoms: Their requirements with regard to source brilliance and coherence, interferometer size and stability, as well as detection efficiency still necessitate the development of new experimental methods for controlling nanoparticles.

In contrast, near-field phenomena, such as the Talbot-Lau effect, allow one to operate with particle beams of modest coherence, without the need for a spatially resolving particle detector, and one can draw on favorable length and mass scaling properties. In order to show this, we start by introducing some elementary coherence considerations and the basic idea behind Talbot interference.

\subsection{Coherence considerations}
In the absence of external forces, the stationary Schr\"odinger equation of quantum mechanics is formally equivalent to the  Helmholtz equation that governs the propagation of light.
This explains why many phenomena from classical wave optics, such as
diffraction and interference, find a close analogy in non-relativistic quantum mechanics. Indeed, Huygens' principle of elementary wavelets and the Kirchhoff-Fresnel integral formula are closely related to a Feynman path integral formulation of the dynamics of matter waves \cite{Storey1994}.

Both in classical optics and in quantum mechanics the ability to observe wave phenomena relies on the preparation of sufficient spatial and temporal coherence, i.e. of stable correlations between separate space-time points of the complex wave field. They should be appreciable over a significant portion of the diffracting element, e.g. over at least two slits of a diffraction grating.

Most matter wave experiments are operated with a particle beam propagating in a well-defined longitudinal direction. In many cases it is then justified to decouple the forward direction from the transverse state of motion.
If we assume an initially incoherent particle source,  its spatial (transverse) coherence depends on the effective width  $a$ of the source aperture.
According to the \emph{van Cittert-Zernike theorem} the spatial coherence behind an incoherent source is described by the same functional form that also quantifies the intensity pattern due to diffraction behind the same aperture under coherent plane wave illumination \cite{Born1993}.
The coherence width at a distance $L$ behind the intrinsically incoherent particle source can therefore be estimated as $2L\ldb/a$. This illustrates that for massive particles with small de Broglie wavelengths $\ldb$ we will need either a narrow source opening $a$ or a long propagation distance $L$ to prepare the required spatial coherence.

Similarly, the \emph{Wiener-Khinchin theorem} describes the longitudinal coherence function as the Fourier transform of the beam spectral density \cite{Born1993}. A narrow distribution of de Broglie wavelengths, i.e. a good momentum selection, is therefore required if we want to prepare longitudinally extended matter wave coherence. This is a source property that cannot be improved by increasing the distance between the source and the grating.

\begin{figure}
  \includegraphics[width=\columnwidth]{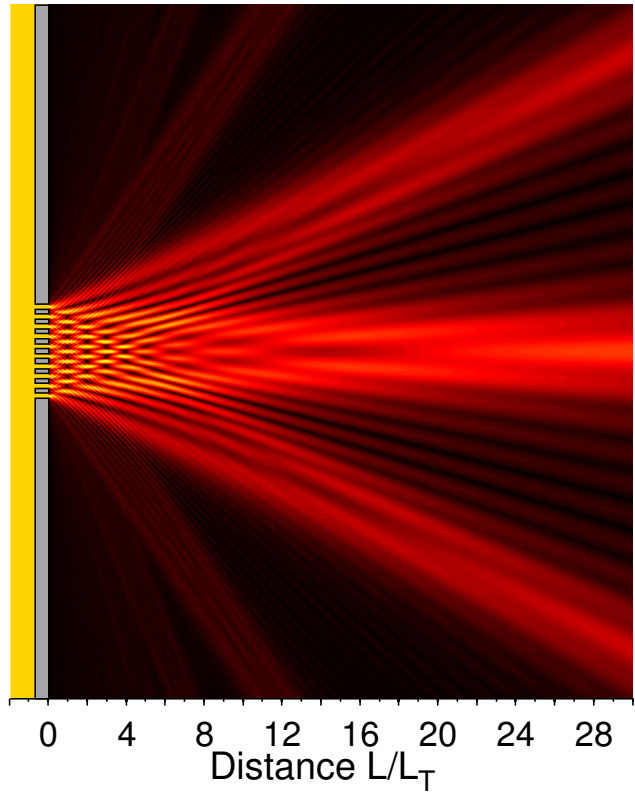} \\
\caption{Transition from near-field to far-field interference for a plane wave illuminating a grating with 10 equidistant slits from the left. The brightness is scaled to be proportional to the wave amplitude. In the near-field, at integer multiples of the Talbot length $L/L_{\rm T}=1,2,3$, one observes partial recurrences of the intensity distribution in the slits, while at greater distances the main diffraction orders of far-field interference start to emerge. 
The calculation is done in the paraxial approximation, assuming ideal gratings with a slit width of half the slit separation. Note that the axes are not to scale.
}
\label{fig:near-far}
\end{figure}

\subsection{Entering the near field}
\label{sec:nearfield}
Many textbooks on classical optics restrict themselves to far-field interference pattern described by the Fourier-transform of the diffraction mask. Also the first diffraction experiments with C$_{60}$ fullerenes were performed in this regime: an effusive molecular beam with a velocity of about $100$\,m/s was sent onto a nanomechanical grating  with a slit separation of  $100\,$nm \cite{Arndt1999}. Given the de Broglie wavelength of 5\,pm, sufficient transverse coherence could only be prepared by reducing the effective source width to smaller than $10\,\mu$m.
Since the first order interference fringes were separated by only $50\,\mu$m at a distance of 1\,m behind the grating, a second $10\,\mu$m slit was placed immediately in front of the grating to collimate the beam width to smaller than the diffraction angle. This concept clearly worked, but at the expense of reducing
the detected particle flux by many orders of magnitude. Experiments with more massive species must cope with even smaller de Broglie wavelengths and even stricter coherence requirements. Several strategies are conceivable to fulfill them.

\begin{figure}
  \includegraphics[width=\columnwidth]{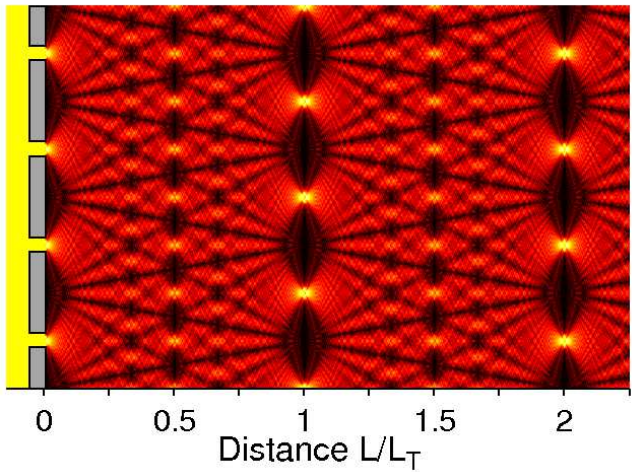} \\
\caption{
Detail of a Talbot carpet in the near-field region of an extended grating generated by a plane wave from the left. Like in Fig.~\ref{fig:near-far} bright colors indicate large wave amplitudes. 
At integer multiples of the Talbot length $L_{\rm T}$ it displays an exact image of the intensity distribution in the grating (shifted by half a grating period for odd multiples).  
To highlight the fractional recurrences at rational multiples of $L/L_{\rm T}$
the calculation was done (in paraxial approximation) with a small slit width of 15\,\% of the grating period.
The Talbot-Lau interferometers described below are based on the same resonant interference effect which gives rise to these structures.
} 
\label{fig:TalbotSimulation}
\end{figure}

Novel source methods may serve to improve the coherence and to increase $\ldb$.
However, slowing of the particle beam alone does not solve the problem, since it reduces only the forward velocity and therefore increases the beam divergence by the same factor.
Genuine cooling would reduce the mean velocity, the velocity spread in all directions, and eventually also the number of internal excitations. But cooling schemes for interferometry with very massive particles are still in an early development stage (see Sect.~\ref{ExperimentalChallenges}).
Improved detectors with true single-particle sensitivity may allow one to compensate for the tiny fluxes associated with a tight collimation.
We focus here on a third aspect, the implementations of novel near-field interferometer schemes which relax the coherence requirements and at the same time parallelize the diffraction effect thousandfold. 

Coherent diffraction at an arbitrary aperture is generally described by the Kirchhoff-Fresnel  integral \cite{Born1993}.  
It can be viewed as a decomposition of the diffraction pattern into spherical waves emanating from all points in the aperture surface. Each contributes a phase $2\pi R/\lambda_{\rm dB}$ with $R$ the distance  between an aperture point and an image point on a screen. 
The paraxial approximation holds if the latter is located at a distance $L$ large compared to the extension of the aperture. The contributing phases can then be expanded to second order in the lateral coordinates of aperture and screen, $\phi = \pi  (x_a-x_s)^2/\ldb L$. 

One can further distinguish between far-field and near-field interference based on the importance of the propagating wavefront curvature in $\phi$  \cite{Born1993,Berman2010}. In the  Fraunhofer, or far-field approximation the quadratic terms $x_a^2$ and $x_s^2$ may be neglected in $\phi$, while they are required for Fresnel, or near-field diffraction. Figure \ref{fig:near-far} shows the transition from the near-field to first features of the far-field regime for diffraction at a grating of ten slits. The proper far-field limit is then reached at the characteristic distance $ a^2/\ldb$, where the wave pattern is already expanded to well beyond the size of the aperture  $a$, so that spherical waves can already be locally approximated  as plane waves.

Our near-field interferometers are based on the principle of {\em coherent lensless self-imaging}. This concept was first developed for light optics \cite{Talbot1836} and it is nowadays often employed  in situations where refractive optical elements are unavailable, such as with molecules \cite{Brezger2002} or x-rays \cite{Pfeiffer2006}.

The effect, which can already be recognized in the near-field region of Fig.~\ref{fig:near-far}, is illustrated in its idealized form in Fig.~\ref{fig:TalbotSimulation}: When a monochromatic plane wave illuminates a wide grating of period $d$, interference of all diffraction orders will reproduce a self-image of the intensity distribution in the mask, at the distance
\begin{equation}
\label{eq:LTdef}
L_{\mathrm{T}}= d^2 / \ldb
\end{equation}
further downstream. The Talbot length  $L_{\mathrm{T}}$ is named  after Henry Fox Talbot who discovered the effect with  light \cite{Talbot1836}. Self-imaging recurs at integer multiples of $L_{\mathrm{T}}$ up to the point where diffraction at the edges of the grating window becomes relevant. At odd multiples of $L_{\mathrm{T}}$  the grating image is shifted by half a fringe period, while it appears unshifted at even multiples; some authors therefore prefer to define $2d^2/\ldb$ as the Talbot length.
Figure~\ref{fig:TalbotSimulation} also reveals fractional revivals with smaller periods in distances which are rational fractions of  $L_{\mathrm{T}}$. The observed intensity distribution is a beautiful example of wave physics and clearly incompatible with the assumption of ray optics \cite{Berry2001,Case2009}.

\begin{figure}
  \includegraphics[width=\columnwidth]{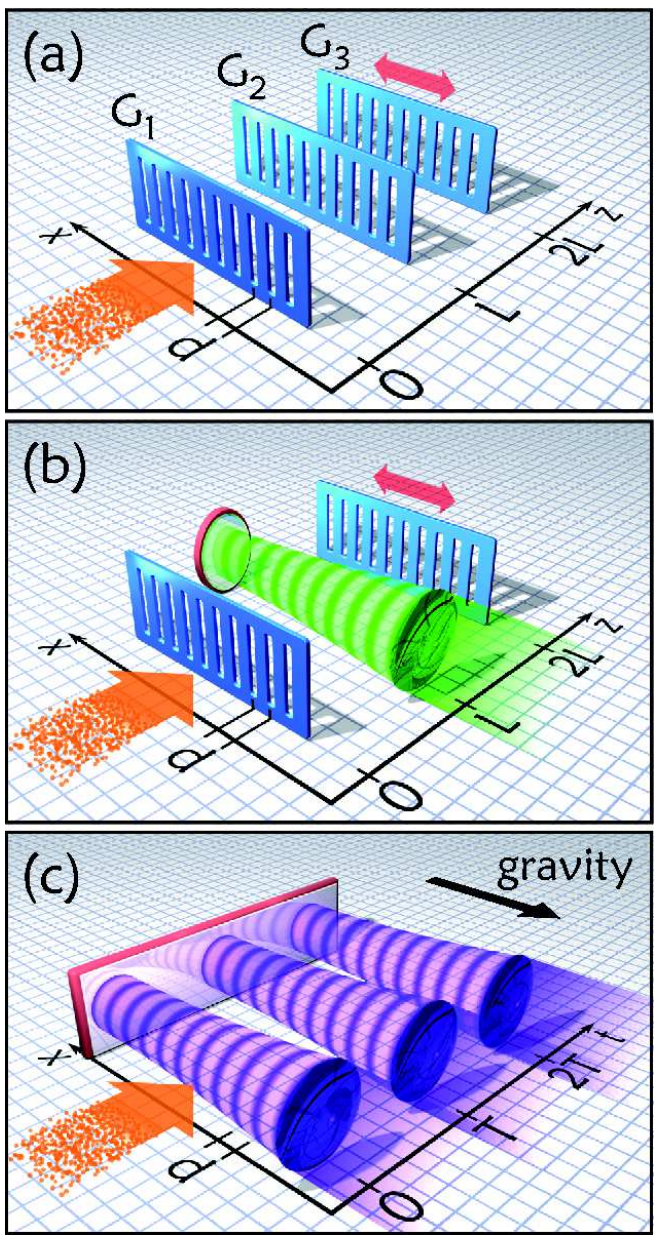}\\
  \caption{Near-field interferometers are optimized for beams of low flux and limited coherence:
  a) The \emph{Talbot-Lau-interferometer (TLI)} uses three nanofabricated gratings. G$_{1}$  prepares the required spatial coherence, G$_{2}$ diffracts the wavelets so that they will form an interferogram, and G$_{3}$ is scanned to sample the fringe pattern  \cite{Clauser1994,Brezger2002}.
  b) The \emph{Kapitza-Dirac-Talbot-Lau interferometer (KDTLI)}  eliminates the dispersive van der Waals interactions between the molecules and the grating walls inside the TLI,  since the central diffraction element is realized as an off-resonant standing light wave \cite{Gerlich2007}.
  c) An \emph{Optical time-domain ionizing matter  (OTIMA) interferometer}, which has yet to be realized, consists of three ultraviolet laser beams. Single-photon ionization in the anti-nodes of the standing wave removes clusters from the incident particle cloud, acting similarly as the massive bars of a material grating. Operated in the time-domain and using short and precisely timed laser pulses, the OTIMA concept avoids many phase averaging effects \cite{Nimmrichter2011}.}
\label{Interferometers}
\end{figure}

\subsection{Concept of Talbot-Lau Interferometry}
The implementation of the Talbot effect still requires a high degree of coherence and has so far been observed with atoms \cite{Chapman1995a,Nowak1997} and electrons \cite{McMorran2009} but not yet with molecules. We can extend this concept to spatially incoherent sources by adding a second grating of the same period.  This leads to the  configuration of array illuminators which  was proposed by \textcite{Lau1948} and is nowadays widely used in light optics \cite{Patorski1989}.

The intuitive picture behind this scheme is the following: The first grating G$_{1}$ acts as a periodic spatial mask which prepares transverse coherence by slicing the incident wave field into numerous wavelets. There is no phase coherence between the waves emerging from neighboring slits. However, each of the individual wavelets emanating from any of the slits of G$_1$ develops sufficient transverse coherence on its way towards the second grating G$_{2}$ to cover two or more slits with a well-defined phase relation. This requires G$_2$ to be at a distance comparable to $L_\mathrm{ T}$ away from G$_1$.
Diffraction at the second grating followed by the free evolution over the Talbot distance then leads to the formation of a spatially periodic intensity pattern whose period equals that of the two gratings. This way, each slit in G$_{1}$  gives rise to a fringe pattern at the detection screen. All interferograms associated with the individual source slits are synchronized in their phase position such that they add up to a high contrast density pattern.
A formal treatment shows that the fringe visibility may actually have a minimum when the grating distance matches exactly the Talbot length \cite{Dubetsky1997a,Nimmrichter2008}, but it reaches a maximum in the close vicinity of $L_{\mathrm{T}}$.

A direct way to visualize the final molecular density pattern is to capture and image the molecules on a clean surface  \cite{Juffmann2009}. Alternatively, one may scan the interferogram with a third grating G$_3$ of the same period \cite{Brezger2002}. In this case the spatial resolution is provided by the grating, so that the integration over a large area leads to a significant gain in signal and a reduction of the measurement time.
This is of particular advantage for sources with a small flux and limited coherence. It also provides a fast signal readout which is often required for feedback and alignment purposes.

In Talbot-Lau interferometry the required grating period $d=\sqrt{\ldb L}\propto m^{-1/2}$  is determined by the de Broglie wavelength and the size of the interferometer $L$, independently of the molecular beam width $D$.
This scaling is much less demanding than that of far-field diffraction,  $d\leq L \ldb/ D \propto m^{-1} D^{-1}$.
In comparison to single-grating far-field diffraction, Talbot-Lau interferometry with three masks imposes more stringent alignment requirements since both the rotation and the longitudinal position of the gratings are important. On the other hand, the arrangement of thousands of parallel slits increases the signal by several orders of magnitude over far-field experiments.

\section{Near-field interference with nanoparticles}\label{sec:interferometers}
Both the short de Broglie wavelength and the limited coherence of available molecular beam sources are the reasons why interferometry with large molecules only started off with the advent of near-field interferometers. The concept was first demonstrated for potassium atoms by \textcite{Clauser1994} and it  was further explored in a number of theoretical  papers \cite{Clauser1997,Dubetsky1997,Rohwedder1999}. Macromolecule interferometry was then realized in  the following steps.

\subsection{Talbot-Lau Interferometry (TLI)}
\label{sec:TLI}
The first fullerene interferometer was implemented in a Talbot-Lau (TL) configuration with three microfabricated gold gratings, as illustrated in Fig.~\ref{Interferometers}a.
The microstructures were written with a period of 991\,nm and an open slit width of about 470\,nm into 16\,mm wide and 500\,nm thin gold  membranes.   Three identical gratings were adjusted with  a separation of 22\,cm \cite{Brezger2002}.

In the experiment care was taken to align all grating slits with an accuracy of about 1\,mrad with respect to each other and to the Earth's gravitational field. The interferometer was placed in a high-vacuum chamber evacuated to better than $10^{-7}$\,mbar and isolated from floor vibrations by an inflated optical table. A sublimation source emitted a thermal beam of molecules, with an internal and motional temperature of up to 900\,K, in the case of C$_{60}$ and C$_{70}$  fullerenes.
\begin{figure}
  \includegraphics[width=\columnwidth]{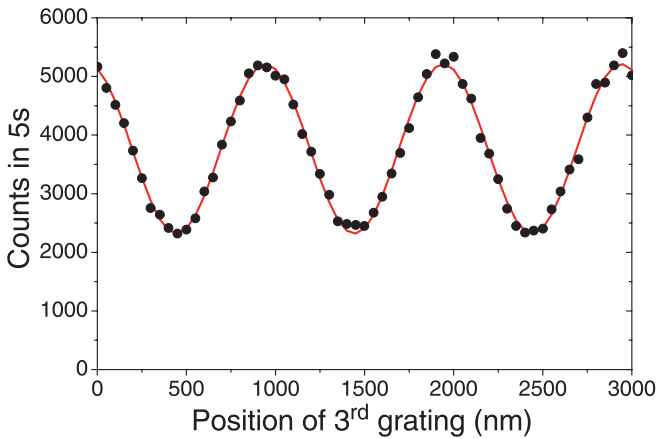} \\
\caption{Talbot-Lau interference of  C$_{70}$ as seen in experiments by \textcite{Brezger2002}. The data set is well reproduced by a sinusoidal curve with a fringe visibility of $38$\,\%.}
\label{C60TLI38percent}
\end{figure}
The thermal distribution contained a broad range of initial velocities.
A velocity band of  $\Delta v/v \simeq 20\%$ was filtered out by selecting the molecular free-fall parabola.
\begin{figure}
  \includegraphics[width=\columnwidth]{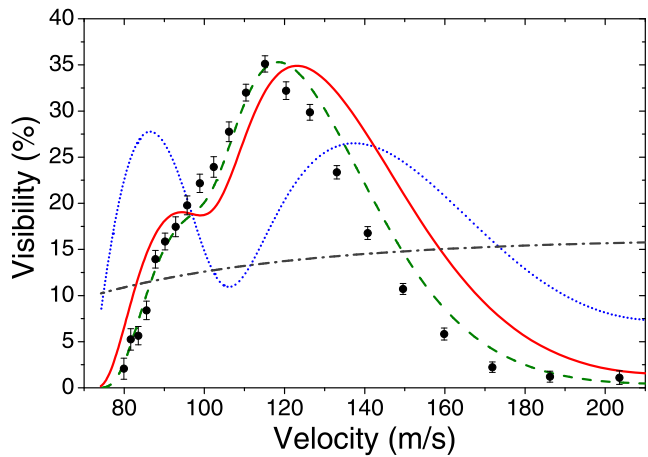}\\
  \caption{Interference in the Talbot Lau configuration leads to near-sinusoidal fringes when the open fraction is approximately 48\,\%. The qualitative and quantitative dependence of the fringe visibility of the molecular velocity is important for discriminating the quantum wave behavior from classical moir{\'e} patterns \cite{Brezger2002}. Full circles: experimental data with statistical error bars. Dotted line: quantum wave model, ignoring the molecular polarizability; Solid line: quantum wave model including the attractive van der Waals interaction between the polarizable C$_{70 }$ molecule and the gold grating wall. Dashed line: Replacing the van der Waals attraction by the asymptotic form of the  Casimir-Polder interaction. 
  Dash-dotted line: Classical shadow contrast in the presence of van der Waals forces.
  All theoretical curves include an average over the measured velocity distribution in the beam.
  }
\label{TLIC70Interference}
\end{figure}
In this interferometer arrangement, high-contrast interference could be observed both for the  fullerenes by \textcite{Brezger2002} and for the biodye tetraphenylporphyrin (TPP) by \textcite{Hackermuller2003}.
In Fig.~\ref{C60TLI38percent}  we show an interferogram of C$_{70}$ displaying high contrast and good signal-to-noise. As in light optics, we define the fringe visibility  of a sine wave pattern as the ratio of its amplitude over its offset.

In contrast to far-field diffraction, which is an unambiguous wave phenomenon without any ray optics analog, the appearance of molecular density fringes in a two-grating experiment might also be explained by a classical moir\'e shadow, to a certain degree.
Fortunately, quantum diffraction and moir\'e effects can be well distinguished by studying the fringe visibility as a function of the particle velocity, as done in  Fig.~\ref{TLIC70Interference}.
If molecules were classical particles travelling along straight trajectories, their velocity would not influence the moir\'e visibility.  They should, therefore, all exhibit a visibility of only $4$\,\% in our settings. In contrast to that, the de Broglie wavelength of a quantum object is  inversely proportional to its velocity and we expect a periodic recurrence of the fringe visibility with the velocity.

Interestingly, this idealized quantum wave picture (dotted line in Fig.~\ref{TLIC70Interference}) does not  reproduce the fullerene interference experiment at all (black circles), since it ignores the van der Waals interaction of the polarizable molecule with the grating walls.
Although the effect of grating interactions was already observed in far-field diffraction at thin SiN masks \cite{Grisenti1999,Nairz2003}, it becomes much more dominant in near-field interference, as discussed in Sect.~\ref{Theory}. 

The C$_{70}$ experiment  can be much better described by a quantum model that includes the van der Waals attraction as in Fig.~\ref{TLIC70Interference}.  The solid line assumes the non-retarded van der Waals potential $V=-C_3/r^3$, with $C_3=$10\,meV\,nm$^{3}$ \cite{Jacob2011}, and the dashed line the asymptotic long-distance form of the Casimir-Polder potential $V=-3\hbar c \alpha_{\mathrm{stat}} / 32 \pi^2\epsilon_0 r^4$, with the static polarizability $\alpha_{\mathrm{stat}} / 4\pi \varepsilon_0 =102\times  10^{-30}\,$m$^3$ for C$_{70}$ \cite{Compagnon2001}. 
The exact potential according to \textcite{Casimir1948} leads to a curve very close to that of the non-retarded form in the present case. An explanation of the remaining discrepancy between experiment and theory requires future experiments with gratings of variable thickness and different materials, as well as better velocity selection.
We note that also the classical treatment must include the attractive force in the grating slit. The expected visibility then follows a monotonic curve in the same diagram (dash-dotted line in Fig.~\ref{TLIC70Interference}), but never exceeds 16\,\%  in our setting.  This is clearly different from both the quantum description and the experimental observation.

The polarizability of a mesoscopic particle is roughly proportional to its volume. 
Therefore Casimir-Polder forces have an even more significant effect on larger particles. This explains why in a Talbot-Lau interferometer with material gratings quantum interference may be difficult to observe for more massive particles,  unless we are either able to prepare a sufficiently intense molecular beam with a narrow velocity spread  below 1\,\% or to fabricate an atomically thin diffraction mask, for instance from graphene \cite{Geim2007}.

\subsection{Kapitza-Dirac-Talbot-Lau Interferometry (KDTLI)}
\label{sec:KDTLI}
\begin{figure}[tb]
  \includegraphics[width=\columnwidth]{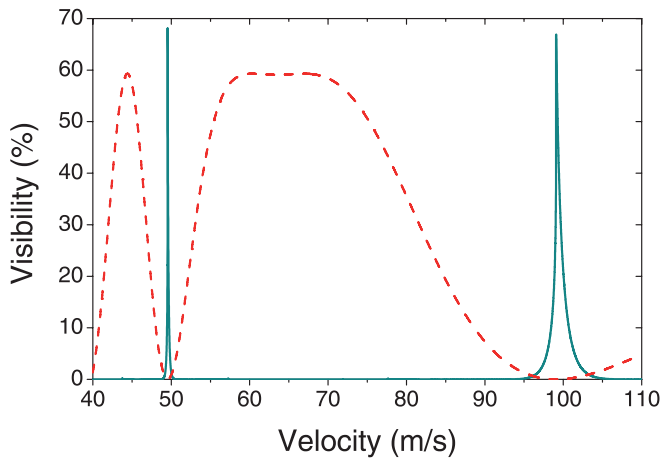}
  \caption{Interference fringe visibility versus molecular velocity for molecules of high polarizability. Dashed line: KDTLI prediction for the perfluoralkylated nanosphere PFNS8 with a polarizability of $\alpha_{\mathrm{opt}}/4\pi \varepsilon_0 =2\times 10^{-28}\,$m$^3$. Solid line: TLI prediction for the same experimental arrangement, except for the laser grating being replaced by a 200\,nm thick SiN wafer with a grating period of 266\,nm and 90\,nm open slits. Even though the use of mechanical gratings may allow one to achieve a high interference contrast at a few selected velocities, real-world sources  have a finite velocity spread and the effective visibility in a TL interferometer can be dramatically reduced when the velocity spread exceeds  1\,\% \cite{Gerlich2007}.}
\label{fig:KDTLIVDependence}
\end{figure}

The problem of van der Waals forces can be eliminated by another interferometer scheme: following the demonstration of an optical phase grating for fullerenes by \textcite{Nairz2001} a new  interferometer was proposed by \textcite{Brezger2003} and implemented by \textcite{Gerlich2007}.
This `Kapitza-Dirac-Talbot-Lau' interferometer (KDTLI) combines the idea of Talbot-Lau (TLI) near-field parallelization with the concept of matter diffraction at a standing light wave  as originally proposed by  \textcite{Kapitza1933} for electrons  and first realized with atoms by \textcite{Moskowitz1983} in the Bragg regime.

\begin{figure*}[tb]
\includegraphics[width=0.8\textwidth]{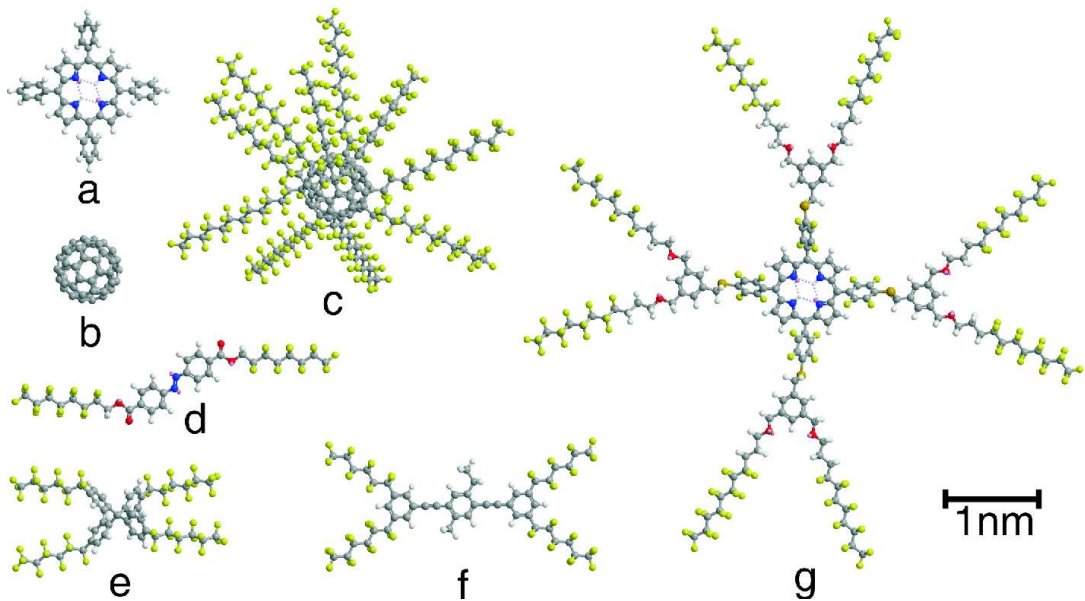}\\
\caption{Gallery of molecules that showed quantum interference in the KDTL interferometer. (a) Tetraphenylporphyrin (TPP); (b) C$_{60 }$ fullerene; (c) PFNS10, a carbon nanosphere with ten perfluroalkyl chains \cite{Gerlich2011}; the variant PFNS8 with eight side arms was also used;  (d) A perfluoroalkyl-functionalized diazobenzene \cite{Gerlich2007}; (e) - (f) two structural isomers with equal chemical composition but different  atomic arrangement \cite{Tuxen2010}; (g) TPPF152, a TPP derivative with 152 fluorine atoms \cite{Gerlich2011}.
}\label{fig:Gallery}
\end{figure*}

The KDTLI evolves from a TLI if we replace the central diffraction grating G$_{2}$ by a standing laser light wave with the same grating period, see Fig.~\ref{Interferometers}. We still keep the mechanical masks G$_{1}$ and G$_{3}$ for coherence preparation  and interference imaging.  The van der Waals phases at these stages do not perturb the final pattern since G$_{1}$ and G$_{3}$ act only as spatially periodic transmission filters (each thinning out the beam by about two thirds).

The new optical phase grating G$_{2}$  does not remove any particles from the beam. Instead, the electric laser field $E$ interacts with the molecular optical polarizability $\alpha_\mathrm{opt}$ to induce a rapidly oscillating electric dipole moment which interacts again with the laser field.
In a standing light wave, the resulting dipole potential $W=- \alpha_\mathrm{opt} E(x,t)^2/2$ is spatially modulated, and so is the imprinted molecular phase. Given that its period is the same as that of G$_{1}$, 
the free evolution of the de Broglie waves behind the grating then results in an observable molecular fringe pattern with the same periodicity.
As illustrated by Fig.~\ref{fig:KDTLIVDependence}, the elimination of the molecule-wall-interaction at G$_{2}$ significantly reduces the monochromaticity requirement on the incident matter waves.

In our experiment the optical phase grating is realized by retro-reflecting a 532\,nm laser of up to 18\,W at a flat mirror. Since the incident molecular beam has a divergence of about 1\,mrad it is important to orient the standing light wave such that no semiclassical molecular trajectory crosses more than a single node or anti-node of the green light-field. In order to meet this condition the laser is focused along the molecular beam axis to a narrow waist  of $w=20\,\mu$m.
The period of the mechanical masks  G$_{1}$ and G$_{3}$ was also carefully matched to that of the laser grating ($266$\,nm) since  already a deviation exceeding $0.05\,$nm would reduce the interference fringe visibility significantly.
\begin{figure}
    \includegraphics[width=\columnwidth]{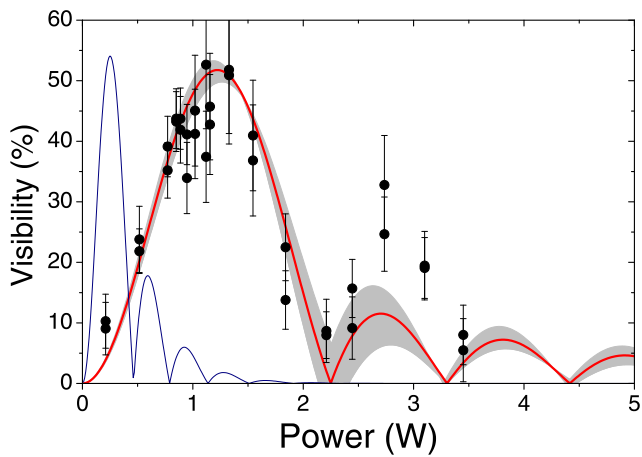}
  \caption{Fringe visibility of PFNS8 in the KDTLI as a function of diffracting laser power at a mean molecular velocity of $75\,$m/s and a velocity spread of $10\%$. Experimental data are represented by full circles, the error bars correspond to a 68$\%$ confidence interval of the sinusoidal fit of the interference pattern. The heavy line is the quantum prediction based on the expected polarizability of $\alpha_{\mathrm{opt}}/4\pi \varepsilon_0 =2\times 10^{-28}\,\mathrm{m}^3$. The shaded area displays the effect of a $\pm 2$\,m/s variation of the mean molecular velocity. The thin line gives the classically expected visibility \cite{Gerlich2011}.}
\label{fig:KDTLIPowerDependence}
\end{figure}

The gratings are mounted on a common base with a mutual distance of 105\,mm. This allows operating with a de Broglie wavelength down to 1\,pm,  for instance with particles of $10,000$\,amu at a velocity of 40\,m/s or correspondingly higher masses at lower velocities.
This design has been experimentally validated in our lab with a variety of organic molecules, as shown in Fig.~\ref{fig:Gallery}. We could demonstrate high contrast quantum interference for many molecular shapes, symmetries, masses and internal excitations. Some of these molecules contained more than 400\,atoms, weighed about 7000\,amu, were thermally excited to 1000\,K  or measured more than $6\,$nm in diameter---and still they all exhibited clear quantum interference \cite{Gerlich2011}.
In Fig.~\ref{fig:KDTLIPowerDependence} we present the evolution of the fringe visibility for the perfluoroalkylated nanosphere PFNS8 (356\,atoms,  5672\,amu) as a function of the diffracting laser power in G$_{2}$. The experimental data (full circles) shows a visibility of up to 50\,\% and it is well described by the quantum predictions (heavy line) and cannot be reproduced by classical physics (thin line).  
The relatively large error bars in this experiment are due to the short measurement time resulting from the thermal instability of the species and from the fact that they are available only in small amounts.

\subsection{Interferometry with pulsed optical gratings (OTIMA)}
\label{sec:OTIMA}
Although KDTL interferometry is compatible with high molecular masses, van der Waals forces in  G$_{1}$ and G$_{3}$ may eventually lead to a blockage of the grating due to adsorbed molecules clogging the slits.
It is therefore useful to consider an all-optical setup, such as the optical time-domain ionizing matter (OTIMA) interferometer we describe below.

As long as we are limited to incoherent sources, the first grating must act as an absorptive mask to prepare the required spatial coherence. Optical amplitude gratings were already realized by \textcite{Abfalterer1997a} for metastable atoms by inducing transitions to undetected states.
Since the high level density in clusters and molecules usually precludes this resonant excitation scheme, \textcite{Reiger2006} proposed photo-ionization gratings as a universal tool for complex nanoparticles, where a single photon suffices to ionize and remove the particles from the anti\-nodes of a standing light wave.  The intensity maxima thus play the role of the grating bars, but their transmission, periodicity, and the additionally imprinted phase can be tuned by varying the pulse energy and the laser wavelength.
In variance to the TLI and KDTLI design, the second pulse G$_{2}$ acts as a combination of both an  absorptive and a phase grating.

The use of pulsed optical gratings also allows one to implement an interferometer in the time-domain as discussed in detail by \textcite{Nimmrichter2011}. Time-domain interferometry was first proposed by \textcite{Moshinski1952} for neutrons. Since then it has been implemented in various atom experiments, for instance by \textcite{Kasevich1991a,Szriftgiser1996,Cahn1997,Fray2004,Turlapov2005}. It permits one to eliminate many velocity dependent dispersive effects since all particles will interact with all perturbations for the same period of time.

\begin{figure}
  \includegraphics[width=\columnwidth]{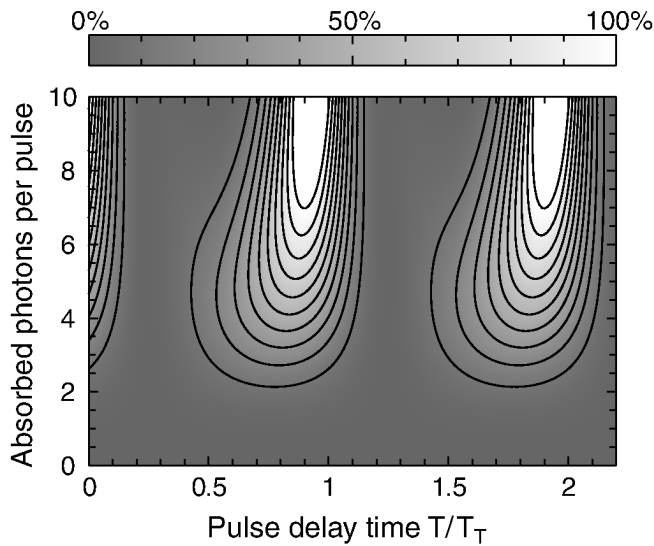}\\
  \caption{Predicted fringe visibility as a function of both the pulse delay time and the mean number of absorbed UV photons. Matter-wave self-imaging occurs at multiples of the Talbot time $T_{\rm T}$. Each single-photon ionization grating leads to a spatially periodic removal of clusters from the beam. The antinodes of the standing light waves thus play the role of the mechanical bars in a nanofabricated grating. By increasing the laser pulse energy one broadens the zones from which the clusters can be efficiently extracted. This corresponds to a reduction of the effective open slit width in a mechanical grating. All gratings are not only ionizing, but are accompanied by an additional phase modulation arising from the interaction between the polarizable molecules and the laser field. This intensity dependent phase gives rise to the predicted asymmetry of this figure.}
\label{VisOtimaQm}
\end{figure}

A possible implementation for clusters is illustrated in Fig.~\ref{Interferometers}c, 
where a  cluster package passes along a plane mirror surface. It is subjected to three retro-reflected UV laser pulses which form standing light waves. These pulses are separated   in time by a pulse delay  of several tens of microseconds for low-mass clusters and up to tens of milliseconds for clusters around 10$^6$\,amu.
The time-domain analog of the Talbot length is the Talbot time $T_{\rm T}=md^{2}/h$, proportional to the particle mass.
For light with a wavelength of $2d=157$\,nm as generated by an F$_2$-excimer laser it amounts to 
$\rm{T}_{\rm T}/\mathrm{ns}=15\,m$/amu.  F$_2$-lasers have a coherence length of about 1\,cm \cite{Sansonetti2001} and a pulse width of about 10\,ns, which suffices to form a laser grating up to a few millimeter distance from the mirror surface. Other light sources, such as high harmonics of solid state lasers, are similarly conceivable.

One may scan the interference fringes for instance by allowing the particles to fall freely in the gravitational field. Since the delay time is the same for all clusters they will fall by the same distance and will thus be effectively scanned by laser grating G$_{3}$, independent of their velocities. The simulation presented in Fig.~\ref{VisOtimaQm} shows that the quantum visibility in such a scheme could reach as much as 100\,\% for a given mass.

\section{Probing environmental decoherence}\label{sec:decoherence}
The experiments described in the previous section demonstrate that complex particles can
be delocalized over hundreds of nanometers, a distance exceeding their size by orders of magnitude. Given this clear confirmation of quantum mechanics, how can one understand that under normal circumstances molecules appear as well-localized objects? This  distinction of states with well-defined position can be explained within the framework of quantum mechanics by the concept of environmental decoherence \cite{Zurek2003,Joos2003,Schlosshauer2007}.

The theory accounts for the crucial influence of practically unobservable environmental degrees of freedom, such as ambient gas particles or the radiation field.
The interaction correlates the environmental quantum state with that of the molecular motion, implying that some information on
the molecule's whereabouts could be obtained in principle by an appropriate measurement of the environment. Even though this cannot be done in practice, the mere fact that which-way information remains in the environment suffices to affect the reduced state of the molecule in the same way as if the particle position was continuously monitored by a coarse-grained detector and the outcome discarded. This leads to the effective localization of the particle, i.e.~to the reduction of spatial coherence, prohibiting its wave behavior in agreement with the complementarity principle \cite{Bohr1949}. Equivalently, one can view the environment as exerting random momentum kicks on the molecule, which blur the molecular state in the momentum representation.

The near-field interference setups discussed in Sect.~\ref{sec:interferometers} are particularly well suited for quantitative decoherence studies since a molecule will typically not be scattered out of the detected beam after an environmental interaction. This enabled the first studies of thermal and collisional decoherence, two paradigmatic mechanisms which can be experimentally well controlled and where the observed reduction of the interference visibility could be compared with the quantitative predictions of decoherence theory.

It should be emphasized that the fringe visibility may also be degraded by more mundane effects which cannot be related to the dissemination of which-way information, even though they may be hard to distinguish from proper decoherence. In practice the most relevant of these is the blurring of the observed interference pattern due to the
phase averaging caused by vibrations and due to thermal drifts of the grating positions and their alignment; also fluctuations of electromagnetic field gradients can reduce the recorded interference visibility.
Such classical noise effects were sufficiently suppressed in the experiments on thermal and collisional decoherence.

\begin{figure}
  \includegraphics[width=\columnwidth]{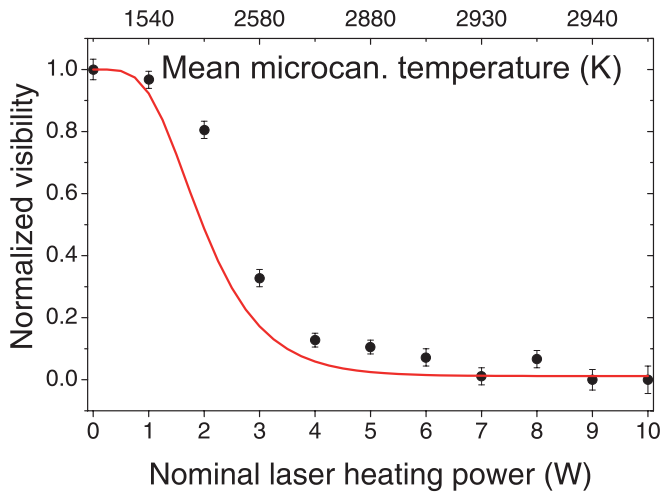}\\
  \caption{Observation of thermal decoherence in a Talbot-Lau interferometer. The expected visibility reduction (solid line) is in good agreement with the experimental observation (points). The bottom scale gives the heating laser power, the top scale shows the mean molecular temperature at the interferometer entrance. The maximal interference path separation of 990\,nm is comparable to the wavelengths of the thermal photons, implying that more than a single photon has to be emitted to fully destroy the fringe visibility.  Combined with the highly nonlinear temperature dependence of the emission probability, this explains the particular form of the curve \cite{Hackermueller2004}. In the experiment the gratings are separated by 38\,cm and the mean beam velocity is 100\,m/s.
\label{ThermalDecoherence}}
\end{figure}

\subsection{Endogenous heat radiation}
Every complex particle with a finite temperature emits thermal radiation. The localization due to that radiation is thus a basic decoherence effect expected to occur in any thermal object. It can be studied conveniently with fullerenes since they behave  in many ways like a small solid when heated to high internal energies.

At temperatures exceeding 1000\,K fullerenes radiate in a continuous optical spectrum, similar to a black body \cite{Hansen1998}. They also start to evaporate C$_{2}$ subunits and to emit thermal electrons. All of these processes can occur while the hot molecules traverse the Talbot-Lau interferometer. However, ionization and fragmentation lead to a complete loss of the molecules and thus do not contribute to the recorded signal.

In the experiment the fullerenes were heated by several intense laser beams in front of the interferometer  \cite{Hackermueller2004}. Both the ionization yield in the excitation region and the increased final detection efficiency are recorded as a function of the heating laser power and the particle velocity, providing a temperature calibration. The agreement of these measurements with a model calculation
yields the distribution of micro-canonical temperatures in the molecular ensemble \cite{Hornberger2005}. Photoemission is the fastest and most efficient cooling process, and a good portion of the internal energy is emitted  before the molecules enter the interferometer. However, it still is probable for the molecules to emit several near-infrared or even visible photons during their transit between the first and the third grating.

The theoretical account of the expected decoherence must consider that fullerenes differ from ideal blackbody emitters. A microscopically realistic description of the spectral emission rate is obtained by including their known frequency-dependent absorption cross section, their finite heat capacity, and the fact that they are not in thermal equilibrium with the radiation field \cite{Hornberger2005}.

As shown in Fig.~\ref{ThermalDecoherence}, the prediction from decoherence theory is well confirmed by the experimental observation: The interference visibility is gradually reduced with increasing molecular temperature until it vanishes completely. The upper scale gives the mean micro-canonical temperature in the molecular beam, showing that at 1500\,K the fullerenes still behave as quantum waves in this experimental arrangement, while  they are indistinguishable from classical particles when close to 3000\,K. The calculation shows that between three and four photons are typically required to reduce the visibility by one half. This is consistent with the emitted wavelength being comparable to the spartial delocalization of the molecular matter waves.

These studies imply that thermal decoherence can turn into a serious obstacle for interferometry with very complex particles. In particular, the effect suffices to explain the localization of truly macroscopic objects, since the critical temperature for the effective quantum-to-classical transition decreases with increasing size \cite{Hornberger2006,Joos2003}. At the same time, thermal decoherence should be avoidable for particles with masses up to $10^9$\,amu by cooling them to their vibrational ground state, i.~e. below 77\,K; at these masses also the vacuum chamber containing the setup needs to be cooled to avoid decoherence due to blackbody radiation \cite{Nimmrichter2011a}.

\begin{figure}
  \includegraphics[width=\columnwidth]{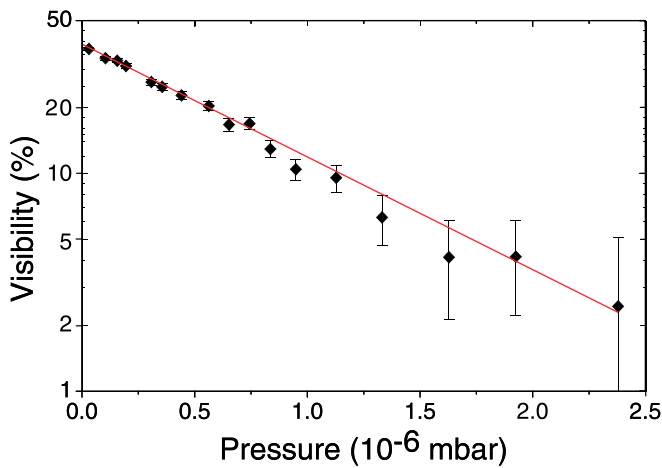}\\
  \caption{
  Interference fringe visibility of C$_{70}$ fullerenes as a function of
  the methane gas pressure in a TLI (semilogarithmic scale). The exponential decay   indicates that already a single collision leads to a complete loss of coherence. Good agreement is found with the prediction of decoherence theory (solid line), which is based on the microscopic scattering dynamics \cite{Hornberger2003a}. \label{CollisionalDecoherence}}
\end{figure}

\subsection{Collisional decoherence}
A second fundamental decoherence effect is related to the
scattering of ambient gas particles off the delocalized molecule.
Using a Talbot-Lau interferometer one can study this effect quantitatively by the gradual admission of different gases into the vacuum chamber. At room temperature, the collisional momentum and information transfer is so high that already a single scattering event per molecule suffices to fully destroy the interference. On the other hand, because of the high mass of the interfering particles and the wide detection area there is no dominant beam depletion due to collisions within the interferometer.

As shown in Fig.~\ref{CollisionalDecoherence} an exponential decay of the fringe visibility can be observed as the gas pressure increases \cite{Hornberger2003a}. This is consistent with assuming that a single collision process is able to resolve the molecular position, or equivalently to blur the interference pattern by the random momentum transfer. We note that also the molecular transmission decreases exponentially, though this is mainly due to collisions outside the interferometer.

We find a good quantitative agreement with decoherence theory, 
as indicated by the solid line in Fig.~\ref{CollisionalDecoherence}.
The calculation is based on a semiclassical approximation for the velocity dependent total scattering cross section, which is determined by the inter-particle van der Waals potential, and which must be averaged over the velocity distribution in the gas, see \textcite{Hackermuller2003}.
This experiment serves also to confirm the short time limit of the quantum linear Boltzmann equation for a tracer particle in a gas \cite{Vacchini2009}.

The interference of  fullerenes yields substantial visibilities even at moderate pressures of $10^{-7}$\,mbar (and an interferometer transit time of 5\,ms to 10\,ms). However, we estimate that quantum interference with gold nanoclusters of $10^{6}$\,amu will require a pressure of less than $10^{-9}$\,mbar \cite{Nimmrichter2011}.

\section{Interference-assisted measurements \label{Metrology}}
\begin{figure}
\includegraphics[width=1\columnwidth]{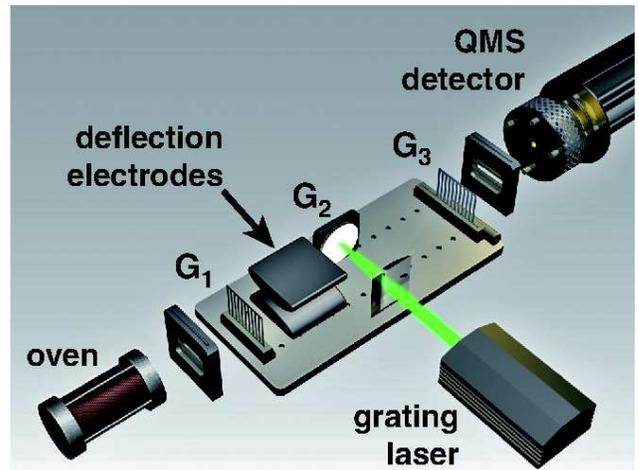}\\
\caption{Interferometric Stark deflectometry combines the high spatial resolution of the KDTL experiment (see Fig.~\ref{Interferometers}b) with the fringe deflection in an electric force field. Beam shifts as small as 10\,nm can easily be resolved and be used to evaluate both intramolecular properties and external forces. From left to right: a thermal source emits an intense beam of neutral molecules. A pair of electrodes, placed in between G$_{1}$ and G$_{2}$, provides the deflection field.  Upper right: electron impact ionization quadrupole mass spectrometer.
}
\label{Deflectometry}
\end{figure}

The narrow spacing of the quantum interference fringes and the high sensitivity of their position to external forces allows one to turn molecule interferometry into a viable tool for quantum-enhanced metrology.
The high potential of interference-assisted measurements has already been demonstrated with atoms. Static polarizabilities \cite{Ekstrom1995,Holmgren2010,Miffre2006A}, the ratio $\hbar/m_\mathrm{Cs}$ \cite{Weiss1993}, the gravitational acceleration \cite{Peters1999}, 
and Earth's rotation \cite{Gustavson1997} have been measured interferometrically, to name a few.

Here we will focus our discussion on the combination of Kapitza-Dirac-Talbot-Lau interferometry with conventional Stark deflectometry \cite{DeHeer2011} to determine the internal properties of large organic molecules. This provides valuable insight for physical chemistry, and it is also important for future interference experiments since the coupling of the molecule to the environment depends on these parameters.

\subsection{Optical polarizability}\label{sec:optical_polarizability}
In KDTL interferometry  the light grating G$_{2}$ interacts with the molecular optical polarizability  $\alpha_{\mathrm{opt}}$ and modulates the de Broglie phase shift. The influence of the laser power $P$ on the fringe visibility $V$ was already illustrated in Fig.~\ref{fig:KDTLIPowerDependence} for the perfluoralkylated nanosphere PFNS8. For smaller molecules with a  higher signal-to-noise ratio, such as the fullerenes, \textcite{Hornberger2009} found that the excellent agreement between theory and experiment permits the determination of  $\alpha_{\mathrm{opt}}$ with an accuracy of around ten percent (dominated by the systematic uncertainties in the power and the vertical waist of the diffracting laser). This allows for instance a clear distinction between C$_{60}$ and C$_{70}$, directly from the power dependence of their fringe visibility.
KDTL interferometry may thus be used for identifying properties  which cannot be discriminated by a mass spectrometer alone \cite{Gerlich2008a}.

\subsection{Electric polarizability}

The KDTLI can be extended by a pair of electrodes \cite{Stefanov2008} in front of G$_{2}$, as shown in Fig.~\ref{Deflectometry}, to access the static polarizabilities.
The inhomogeneous electric field $\textbf{E}$ produced by the electrodes provides a homogenous transverse force on a polarizable particle of polarizability $\alpha_{\mathrm{stat}}$. The magnitude of the final quantum fringe shift
\begin{equation}\label{deflection}
\Delta x=K\frac{ \alpha_{\mathrm{stat}}}{2mv^2}\,\frac{\partial}{\partial x}(\mathbf{E}^2)
\end{equation}
is identical to the displacement of a classical particle beam of mass $m$ and velocity $v$. $K$ is a constant defined by the geometry of the electrodes.  Matter wave interferometry, however, additionally imprints a nanosized, high-contrast fringe pattern whose shift can be monitored with a resolution of a few nanometers.  This exceeds the spatial resolution of typical classical experiments by orders of magnitude. Because of that, quantum experiments can operate at lower fields and be less intrusive.
At present, the experimental uncertainty in quantum deflectometry is still comparable to its classical analogue \cite{Berninger2007}, since the experimental precision is currently limited by the uncertainty of the velocity measurement and the error of the geometry factor $K$.  Future experiments featuring an improved velocity selection and a better calibration, for instance using an atom beam, are expected to  achieve an accuracy of better than 1\,\%. This is close to the intrinsic limit for thermally excited  molecules undergoing state changes on a rapid time scale, as shown below.

\subsection{Structural isomers}
\label{Isomers}
The molecular polarizability is also closely linked to the intramolecular atomic arrangement. Stark deflectometry therefore allows one to distinguish structural isomers, i.e. molecules which have the same chemical composition and mass but a different atomic arrangement. This is the case for compounds (e) and (f) in Fig.~\ref{fig:Gallery} which were tailor-made by \textcite{Tuxen2010}. Molecular beams of both species are prepared under equal conditions and delocalized over similar areas in the same interferometer. The stretched compound  (f), however, contains a delocalized $\pi$-electron system which enhances the electronic polarizability with respect to the value of the compound (e) with a tetrahedral core.
Interferometric deflectometry can then distinguish between these two species and in principle even sort them spatially \cite{Ulbricht2008}.  This occurs without the exchange of any which-path information, since the interaction of the molecules with the external field is conservative and reversible.

\subsection{Thermal dynamics}
In contrast to atoms, whose interaction with an electric field is determined by the static atomic polarizability alone, complex molecules are bestowed with many additional degrees of freedom. Even intrinsically non-polar particles, with a point-symmetric thermal ground state, may develop vibrationally induced electric dipole moments
which fluctuate on a short timescale. The beam shift in an external field is then again determined by a polarizability as in Eq.~(\ref{deflection}). However, in place of only the static polarizability, which describes the response of the electron cloud,  we must use the sum of $\alpha_{\mathrm{stat}}$ and a temperature dependent nuclear term accounting for the nuclear motion. It is determined by the thermal average  $\langle {d}^2\rangle$  of the squared electric dipole component \cite{Compagnon2002}:
\begin{equation}\label{susceptibility}
\alpha_\mathrm{tot} = \alpha_{\mathrm{stat}} + \frac{\langle  {d}^2\rangle}{3k_{\rm B}T}
\end{equation}
Although the underlying conformation changes occur on a subnanosecond time scale and average the  expectation value of the dipole moment to zero, the squared dipole remains finite at finite temperatures.
This picture was experimentally tested with hot perfluoroalkylated azobenzene molecules, shown in Fig.~\ref{fig:Gallery}d. At 500\,K they undergo rapid fluctuations with dipole variations between 0.8 to 3.6\,Debye. Interferometric deflectometry yields a total polarizability of $\alpha_\mathrm{tot}/4\pi \varepsilon_0= (95\pm 3\pm 8) \times  10^{-30}\,\mathrm{m}^3$, which is larger than the value for the electronic polarizability $\alpha_{\mathrm{stat}}/4\pi \varepsilon_0= (61\pm 1 \pm 7) \times  10^{-30}\,$m$^3$, as taken from a visibility-versus-power curve. Here the first uncertainty values give the statistical error, and the second ones the systematic error. The observed total polarizability is numerically consistent with the sum of the electronic polarizability and the temperature dependent term  in Eq.~(\ref{susceptibility})  \cite{Gring2010}.

\subsection{Permanent electric dipole moments}
While all measurements described so far are contrast-preserving at a fixed molecular velocity, the interaction between a static molecular moment and an external field may eventually lead to a loss of the fringe visibility due to the molecular rotation. If a polar molecule is exposed to an external field gradient, a deflection force $\mathbf{F}=-\boldsymbol{\nabla}(\mathbf{d}\cdot\textbf{E})$ will displace the interference pattern in dependence of the molecular orientation with respect to the field. Thermally excited molecules will generally leave the source with a random orientation, rotating at frequencies around $10^{10}$\,Hz. The fringe shift then varies not only in magnitude but also in direction such that the interference pattern washes out already at moderate external field strengths. This is a typical example where the fringe visibility is reduced by phase averaging instead of genuine decoherence, i.e.~without the dissemination of which-path information into the environment \cite{Eibenberger2011}.

\subsection{Absolute absorption cross sections}
The experiments of Sect.~\ref{sec:decoherence} have illustrated how the emission of photons can lead to the decoherence of matter waves.  This sensitivity to the recoil of a single photon can be brought to practical use
in an interferometric measurement of the absolute molecular photoabsorption cross section \cite{Nimmrichter2008a}.
Photon absorption may be induced by the running wave of an additional laser beam crossing the molecular beam in the direction of the grating vector.
Given the Poissonian photon number statistics, a small discrete number of photons can get absorbed, and the corresponding interferograms are shifted in discrete steps which add up to an interference pattern with reduced visibility. By a judicious choice of the laser position the interference visibility is maximally blurred already if a single photon is absorbed on average. It is thus possible
to determine  the absorption cross section  without knowing the molecular density in the beam, a frequent challenge in physical chemistry.
This exemplifies that quantum decoherence phenomena can become a tool for molecular metrology.

\section{Experimental challenges}
\label{ExperimentalChallenges}
Interference experiments are only as good as their beam sources and detectors. In the following, we will therefore briefly describe some schemes which are currently available or under investigation for matter wave interferometry.
For brevity we will limit the discussion to neutral clusters and molecules.

\subsection{Molecular beam sources}

An ideal source emits velocity- and mass-selected particles at low internal and external temperatures into a well-defined direction in ultrahigh vacuum.
Up to now, molecular interference experiments had to rely on either supersonic sources or thermal beams. Effusive sources are particularly appealing for matter wave experiments as they are fully vacuum compatible, simple, and have the capability of generating beams of molecules exceeding 10,000 amu with velocity distributions that have significant components at velocities as low as 10\,m/s. Supersonic sources, on the other hand, generate beams which emerge at much higher velocities, but exhibit much smaller velocity spreads and internal temperatures \cite{Scoles1988}, which makes them still suitable for interferometric metrology experiments with polypeptides and oligonucleotides.
Mechanical slowing mechanisms  \cite{Gupta1999,Narevicius2007} and electromagnetic slowing mechanisms  \cite{Bethlem1999,Fulton2004,Narevicius2008} have been used to decelerate molecules around 100\,amu.  Molecules beyond the size of a single virus may be volatilized by matrix assisted laser desorption (MALDI) \cite{Tanaka1988} or electrospray ionization (ESI) \cite{Fenn1989}.

Large clusters of metals and semiconductors can be prepared using aggregation sources  \cite{Martin1984,Haberland1991,Issendorff2005} with diameters up to several nanometers,  both as neutrals as well as ions,  and with a base velocity of about 200 to 300\,m/s in a carrier gas at 77\,K. Further slowing and internal state cooling can be achieved using a cold buffer gas or cryogenic ion traps.

Low molecular beam velocities are expected via sympathetic cooling by laser cooled ions \cite{Molhave2000} or by cooling in an off-resonant cavity \cite{Horak1997,Nimmrichter2010,Chang2010,Romero-Isart2011}. 
However, many experimental challenges still have to be overcome to turn any of these sources into a reliable method for nanoparticle quantum optics.

\subsection{Detection methods}
All experiments described in this review were carried out using either thermal laser ionization or electron impact quadrupole mass spectrometry.
Thermionic emission of electrons is an efficient and fast detection method for stable particles, such as fullerenes, which exhibit thermal photon emission rather than fragmentation when heated to high temperatures \cite{Campbell1991,Hansen1998}.

Electron impact ionization quadrupole mass spectrometry, on the other hand, is more universal at low masses, but limited to a typical detection efficiency of $ 10^{-4}$ and often compromised by uncontrolled fragmentation at high mass.
Complementary to that, single-photon or two-photon methods seem to be adapted for the detection of clusters of almost any size, yet they often fail for large organic molecules \cite{Hanley2009}.

This is why scanning tunneling microscopy \cite{Juffmann2009} or fluorescence methods \cite{Stibor2005a} have been established to image surface deposited  interferograms. Their high detection efficiency, however, has to be weighed against the difficulty of distinguishing molecules from their fragments.

\subsection{External perturbations}
Classical noise phenomena, which should be distinguished from the decoherence processes described in Sect.~\ref{sec:decoherence}, are conceptually less intriguing, but are in practice often more relevant to the experiment. A theoretical discussion in the context of atom interferometry was given by \textcite{Schmiedmayer1997a,Miffre2006B}, and for molecule experiments by \textcite{Stibor2005}

In a three-grating interferometer the fringe shift $\Delta x$ depends on the relative position of all gratings as given by $\Delta x = \Delta x_1-2\Delta x_2+\Delta x_3$. Grating vibrations may  destroy the interference  fringes already at amplitudes as small as a few nanometers. In the presence of a constant  acceleration $a$ the fringe shift  is $\Delta x = -2(aT^2/2) +a(2T)^2/2=aT^2$, with  $T$ the  free evolution time between two subsequent gratings.
This applies to the Earth's gravitational acceleration $a_g=9.81$\,m/s$^2$, the Coriolis acceleration $a_c=2 \mathbf{v}\times \mathbf{\Omega}_E$, or any constant electromagnetic acceleration.

An overall fringe shift will not destroy the interference pattern. However, if different particles experience different shifts due to  different transit times the final molecular pattern will be a mixture of differently shifted interferograms, and the visibility can be drastically reduced.
An interferometer in the time-domain (see Sect.~\ref{sec:OTIMA}) can eliminate all phase shifts that depend on transit times.
Even the intrinsically velocity-dependent Coriolis force can be compensated then by a reorientation of the interferometer grating vector parallel to $\mathbf{\Omega}_E$.

\section{Theory of Talbot-Lau near-field interference \label{Theory}}
We now provide an overview of the theory, which
permitted us to design the interferometers such that the quantum mechanically predicted fringe visibility is always considerably larger than of the moir{\'e}-type shadows expected by classical physics.

In Talbot-Lau near-field interference, many different diffraction orders contribute to a resonant interference effect. Even tiny distortions of the various wave fronts can therefore result in large effects. This applies in particular to the influence of the dispersion forces between the polarizable molecules and the grating walls. An accurate description must also account for the finite longitudinal coherence in the initial beam, as well as the incoherent effects of photon absorption and Rayleigh scattering in the standing laser light wave. For precise predictions also the finite width and the transverse coherence of the molecular beam entering the interferometer must be included, as well as the effects of grating vibrations and inertial forces due to gravity and the rotation of the Earth.

All this  can be accounted for in a transparent and largely analytical fashion by expressing the state of the particle beam and its evolution in phase space in terms of the Wigner function \cite{Wigner1932,Schleich2001}. This also facilitates the incorporation of  decoherence effects caused by the emission of thermal radiation or the scattering of particles.
The comparison with the classical prediction, including all forces and environmental effects,
then simply requires one to replace the Wigner function by the classical phase space distribution.

\subsection{Phase-space formulation}
A quantum phase space theory of Talbot-Lau interferometry was developed by \textcite{Hornberger2004} and refined to the treatment of grating dispersion forces beyond the eikonal approximation by \textcite{Nimmrichter2008}.
It is based on earlier treatments using wave functions \cite{Patorski1989,Clauser1992,Berman1997,Brezger2003}.
The concept was later extended to Kapitza-Dirac-Talbot-Lau interferometry \cite{Hornberger2009} and to time-domain interferometry with ionizing laser beams  \cite{Nimmrichter2011}.

We assume a coarsely collimated molecular beam where the longitudinal speed exceeds the transversal velocity.
The change of the longitudinal velocity component $v_z$ may then be neglected as the particles pass the interferometer, and the description can be confined to the transverse state of motion in a longitudinally comoving frame.
One considers how the transverse Wigner function transforms
under the sequential steps of passages through the gratings and the stretches of free propagations,
and includes the longitudinal coherence only in the end by averaging
over the velocity distribution in the beam. The classical prediction can be obtained in much the same way,
since the Wigner function and the classical phase space distribution exhibit the same shearing transformation during the free motion in between the gratings.

To keep the presentation simple we focus here on the results for 
the special case of equidistant gratings with equal period $d$, and we assume  a transversally extended and incoherent initial beam.
The periodic nature of the diffraction masks then allows us to expand the expected periodic fringe pattern in a Fourier series,
\begin{equation}
\label{eq:wTLI}
w_\mathrm{TL}(x)=\sum_{m=-\infty}^\infty B_m^*(0) B_{2m}\left(m\frac{L}{L_{\rm T}}\right)\exp\left(2\pi i m \frac{x}{d}\right).
\end{equation}
It involves the Talbot-Lau coefficients
\begin{equation}
\label{eq:TLcoeff}
B_m(\xi)=\sum_{j=-\infty}^\infty b_j b^*_{j-m}\exp[i\pi (m-2j)\xi].
\end{equation}
The $b_j$ components are the Fourier coefficients of the transmission function of the second grating G$_2$,
\begin{eqnarray}
\label{eq:tofx}
t(x)&=&b(x) \exp\left(
\frac{-i}{\hbar v_z}\int V(x,z) \,{\rm d} z
\right)
\nonumber\\
&=&\sum_{j=-\infty}^\infty b_j\exp\left(2\pi ij\frac{x}{d}\right).
\end{eqnarray}
They are determined by the aperture amplitude $0\leq b(x) \leq 1$, and they also involve a  complex phase if the grating potential $V(x,z)$ caused by dispersion forces or optical dipole forces is included.

It follows from Eq.~(\ref{eq:TLcoeff}) that for integer $\xi=n$ the $B_m(n)$ reduce to the Fourier coefficients of the transmission probability $|t(x)|^2$ (shifted by half a grating period for odd $n$). This is the same self-imaging phenomenon encountered if a plane wave illuminates a single grating, see Sect.~\ref{sec:nearfield}. 
Indeed, the density pattern of the  basic Talbot effect consists of the same coefficients (\ref{eq:TLcoeff}), and reads as \cite{Nimmrichter2008}
\begin{equation}
\label{eq:TalbotEffect}
w_\mathrm{T}(x)=\sum_{m=-\infty}^\infty B_m\left(m\frac{L}{L_{\rm T}}\right)\exp\left(2\pi i m \frac{x}{d}\right).
\end{equation}
At integer multiples of the Talbot length, i.e.~$L=n L_{\rm T}$, one indeed recovers the grating profile $|t(x)|^2=|b(x)|^2$. At fractional multiples, $L/L_{\rm T}=n/m$, smaller periods appear in the interferogram. This can be clearly seen, e.~g.~for $n/m=1/2, 1/3, 1/4$, in the carpet of Fig.~\ref{fig:TalbotSimulation}, which was produced with this formula for $V=0$.

The comparison with Eq.~(\ref{eq:wTLI}) shows that in an incoherently illuminated TLI or KDTLI the density pattern $w_\mathrm{TL}(x)$ at the position of the third grating is given by a convolution of the Talbot pattern (\ref{eq:TalbotEffect}) with the first grating mask  $|t(x)|^2$.
If a third grating is used to scan the interferogram, another convolution of (\ref{eq:wTLI}) with the transmission probability $|t(x)|^2$ produces the same form as (\ref{eq:wTLI}) with another factor of $B_m^*(0)$, such that  the transmitted signal has the Fourier components $S_m=  [B_m^*(0)]^2 B_{2m}(mL/L_{\rm T})$.  The fringe visibility of a sinusoidal fit to the density pattern, as done in the experiment, is then obtained as the ratio of the first and zeroth Fourier components, ${\cal V}_{\rm sin}=2 |S_1/S_0|$. The general form of the visibility curves of Figs.~\ref{TLIC70Interference}, \ref{fig:KDTLIVDependence}, and \ref{fig:KDTLIPowerDependence} can be reproduced with these formulas.

The time-domain interferometer of Sect.~\ref{sec:OTIMA} can be described by the same formalism, if we replace the longitudinal position of the comoving frame of reference by the evolved time. The length ratio of $L/L_{\rm T}$ is then replaced by the time ratio $T/T_{\rm T}$.

\subsection{Incorporating decoherence}
Our phase space formulation also allows one to easily incorporate the effects of decoherence discussed in Sect.~\ref{sec:decoherence}.
During the free propagation of the matter wave any scattering or emission event will reduce the off-diagonal elements of the motional density matrix in position representation, $\langle \mathbf{x}|\rho|\mathbf{x}'\rangle \to \eta(\mathbf{x}-\mathbf{x}') \langle \mathbf{x}|\rho|\mathbf{x}'\rangle $. The  decoherence function $\eta$ describes the reduction of the fringe visibility, satisfying $|\eta|\leq1$ and $\eta(0)=1$. It can be calculated for various decoherence processes based on their detailed microscopic physics  \cite{Hornberger2004}.

Given the rate $R(t)$ of interaction events with the environment, the effect of decoherence is accounted for by replacing the Talbot-Lau coefficients $B_m(\xi)$ in (\ref{eq:wTLI}) by
\begin{equation}
  B_m(\xi)
\exp\left(
-\int_{-L/v_z}^{L/v_z}
R(t) \left[
1-\eta\left(
\frac{m d}{2} \frac{|v_z t|-L}{L_{\rm T}}
\right)
\right]
{\rm d}t
\right).
\nonumber
\end{equation}
The $m=0$ coefficient, describing the mean particle current, remains unaffected. The $m=2$ coefficient, on the other hand, which determines the sinusoidal visibility gets reduced by an exponential factor. It is determined by an integral over the decoherence function, whose argument is the effective separation between two neighboring interference paths. We observe that decoherence is most effective at the position $v_z t=0$ of the central grating, whereas there is no effect at $v_z t=-L$ and $v_z t=L$ where all interference paths coalesce. This agrees with the intuitive picture that decoherence is related to the degree of information gained by the environment in the interaction process: Which-path information is best available where the interference paths are farthest apart, i.~e. at the central grating.

In the case of decoherence due to collisions between delocalized molecules and residual gas particles, the decoherence function $\eta$ is given by an angular integration involving the scattering amplitude $f$, the total cross section $\sigma$, and an average over the distribution of gas velocities $v_{\rm g}$ \cite{Hornberger2004},
\begin{eqnarray}
\label{eq:eta}
\eta(x)&=&
\left\langle
\int{\rm d}\Omega
\frac{|f\big(\cos(\theta)\big)|^2}{\sigma(v_{\rm g})}
{\rm sinc}\Big(\sin\Big(\frac{\theta}{2}\Big)\frac{2  v_{\rm g}m_{\rm g}x}{\hbar} \Big)
\right\rangle_{v_{\rm g}}.
\nonumber
\end{eqnarray}
The argument of the
function $\operatorname{sinc}\left( z\right) =\sin\left(z\right) /z$
compares the distance $x$ between the interference paths to the  wavelength associated to the momentum exchange experienced if the gas particle scatters with angle $\theta$.
Thus, the better the probing gas particles can resolve the separation between the interference paths, the stronger the reduction of the fringe visibility. This is analogous to the case of a Heisenberg microscope, where the spatial resolution is determined by the wavelength of the probe particles.

\section{Exploring new physics with mesoscopic matter waves}
\label{FundamentalLimits}
\begin{figure}
  \includegraphics[width=\columnwidth]{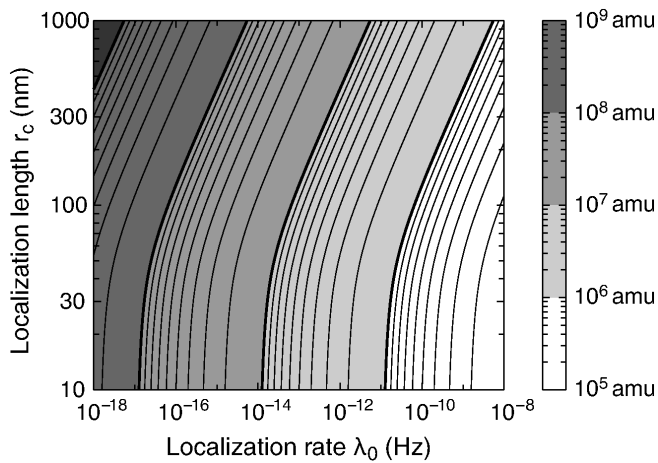}\\
  \caption{Critical mass for testing the Continuous Spontaneous Localization model \cite{Ghirardi1990} in the OTIMA interferometer \cite{Nimmrichter2011} with gold clusters. For a mass beyond 10$^6$\,amu the experiments would already rule out a significant value range of the localization parameters $\lambda$ and $r_{\rm c}$ of the model. High-contrast interferometry with m=10$^8$\,amu would largely exclude the validity of all current estimates of the CSL model.}
\label{fig:csl}
\end{figure}

So far, we discussed the recent progress in high mass interferometry, as well as the challenges related to extensions into the mass range beyond 10$^6$\,amu. Complementary to that, various theories have been put forward over the last decades for an objective modification of the quantum superposition principle for macroscopic objects.
Such speculations range from nonlinear extensions of the Schr{\"o}dinger equation \cite{Bialynickibirula1976,Grossardt2011}, dephasing due to space-time fluctuations \cite{Wang2006}, and  gravitational collapse models \cite{Karolyhazy1966,Diosi1989,Penrose1996}, to spontaneous localization theories \cite{Ghirardi1986,Pearle1989,Ghirardi1990}. These models have in common that they modify the motional dynamics of quantum objects in such a way that the quantum superposition principle fails above a certain mass scale of the objects. This way they lead to a macrorealist description \cite{Leggett2002}, where non-classical delocalized quantum states of macroscopic objects are excluded.

While most of the suggested models provide at most rough estimates of the critical mass range, the theory of Continuous Spontaneous Localization (CSL) by \textcite{Ghirardi1990} has been extensively studied \cite{Bassi2003} and yields quantitative predictions. In the CSL model a stochastic term is added to the many-particle Schr{\"o}dinger equation which randomly collapses the wavefunction to a length scale given by the localization length parameter, commonly estimated as $r_{\rm c} = 100\,$nm. The rate of the localization events grows quadratically with the mass of the composite object. In the CSL literature the rate parameter $\lambda_0$ is specified as the localization rate for a single nucleon ($m=1\,$amu). Current estimates of the strength of the CSL effect by \textcite{Adler2007} and \textcite{Bassi2010} locate its value between $10^{-12}\,$Hz and $10^{-8}\,$Hz, which also ensures that the CSL predictions are consistent with all currently known micro- and mesoscopic quantum phenomena.

\textcite{Nimmrichter2011} have shown that the OTIMA interference experiment outlined in Sect.~\ref{sec:OTIMA} should be able to test the predictions of CSL with nanoparticles in the mass range between $10^6\,$amu and $10^8\,$amu. This is illustrated in Fig.~\ref{fig:csl}, where the critical mass for testing the continuous spontaneous localization is plotted for reasonable values of the free localization parameters $\lambda_0$ and $r_{\rm c}$. Observing interference at $10^8\,$amu in this setting would largely rule out the CSL model in its currently estimated strength.

\section{Conclusions}\label{Conclusions}
Matter wave interferometry with nanoscale objects is still a young  discipline at the interface between the foundations of quantum physics, atomic, molecular and cluster physics, and the nanosciences. We have seen that the de Broglie wave well
describes the center-of-mass motion of even very complex particles, 
giving rise to interference phenomena which can be surprisingly robust against a large variety of internal state transformations and against interactions between external force fields and internal particle dynamics.

In the coming years we expect to exploit the finesse of quantum effects for measuring electromagnetic and structural properties of nanosized objects with growing sensitivity. 
It is important to do so, not only to obtain insights about nanoparticles, but also to assess the feasibility of quantum experiments with ever more complex compounds.
Future explorations should also study the effect of atomic or molecular adducts to the interfering nanoparticles, and the role of thermal, optical, or magnetic properties of the diffracted species.

We have also indicated the interesting prospects for matter wave interferometry with particles in the mass range of $10^6$\,amu and beyond. Various experimental challenges are still to be overcome to get there, but they will enable new tests of decoherence mechanisms and experimental explorations  of standing hypotheses on modifications of established quantum physics.

Cluster interferometry may be contrasted with the enormous progress seen in the
development of ultra-cold Bose-Einstein condensates (BEC), which have opened a new class of atomic  quantum coherence experiments \cite{Anderson1995,Davis1995}.  BEC experiments are complementary to the studies described here.

A single BEC may contain as few as several hundred atoms, comparable to the PFNS10 or TPPF152 molecules, or up to one billion  atoms all occupying a single-particle state. But the parameter range of highly diluted, weakly bound ultra-cold atoms at temperatures below $1\,\mu$K differs by many orders of magnitude from  that of molecular and cluster physics, where hundreds of atoms are bound together in a small but dense piece of condensed matter at internal excitations well above room temperature. In contrast to cold atom experiments, in cluster interferometry the entire compound is delocalized and interferes as a single entity. This is implied by the fact that, unlike in a BEC, the  de Broglie wavelength
is given by the mass of the whole object.

Our experiments are also complementary to proposals for
using mechanical oscillators to test the limits of the quantum superposition principle \cite{Marshall2003, Romero-Isart2011b,OConnel2010}. Mechanical devices are orders of magnitude more massive than even the largest clusters conceivable in foreseeable matter wave experiments. However,  the high cantilever mass limits the maximal spatial separation between two superposed center-of-mass wave packets. It will remain many orders of magnitude smaller than the separation routinely achieved in molecule interferometry.

Bose-Einstein condensation, interferometry with nanoparticles, and quantum studies with nanomechanical oscillators are therefore truly
complementary approaches  to  investigate the nature of macroscopic quantum physics.

\section*{Acknowledgments}
We are indebted to many students, postdocs  and collaborators in Vienna over many years. Special thanks go to Anton Zeilinger for initiating and stimulating the research program on macromolecule interferometry and for continuous encouragement along the way over many years. We thank Marcel Mayor and Jens T\"uxen, University of Basel, for the synthesis and characterization of many molecules described in this work.
We acknowledge financial support  through the Austrian FWF Wittgenstein grant (Z149-N16) and CoQuS (C33N16), as well as the ESF EuroQuasar project MIME (I146-N16).

\end{document}